# Web View: A Measurement Platform for Depicting Web Browsing Performance and Delivery

Antoine Saverimoutou, Bertrand Mathieu, Sandrine Vaton

*Abstract*— Web browsing is the main Internet Service and every customer wants the minimum web page loading times. To satisfy this perpetual need, web browsers offer new processing engines and increased functionalities. Web developers make use of new programming languages and paradigms, new transport protocols have been introduced, network operators offer increased bandwidth and Content Delivery Networks (CDN) providers deploy web servers closest to end-users. To assess the efficiency of these evolutions, delivered performance is often evaluated into a standalone manner, without considering the whole chain of web browsing. We propose Web View, a measurement platform, which performs automated web browsing sessions on popular websites in a user-representative environment. Through different configurable parameters, Web View measures a wide set of information in order to evaluate the impact of the different parameters (transport protocols, web browsers, CDN, access networks, etc.). Based on those measurements, to ease the understanding of web browsing and the corresponding evolutions, Web View offers a public visualization website (https://webview.orange.com). For instance, with Web View, we were able to detect that the use of different CDNs at different times of the day can decrease loading times up to 400% and that the delivery of content through different transport protocols can decrease loading times up to 79%. This paper presents the Web View measurement platform as well as remarkable events we noticed during more than one year of measurements.

## I. INTRODUCTION

The Web was originally meant to deliver static contents but has evolved considerably over the last years to deliver a wide range of contents through different service providers (analytics, advertising, social networking and CDNs). To cope with the increasing usage of dynamic contents embedded in today's web pages, web browser's engines are regularly improves and offer new functionalities. New Internet protocols are defined to improve content delivery such as HTTP/2 (Version 2 of the HyperText Protocol) [1] which runs over TCP and QUIC (Quick UDP Internet Connections) [2] running over UDP, to be standardized as HTTP/3. Network operators offer increased throughput for their access networks and content service providers increase the availability of CDN infrastructures.

One important aspect for Web browsing is the web pages' loading times and recent studies show that high loading times (more than 5 seconds) may lead to web page abandonment. In order to measure web pages loading times, standardization bodies like the W3C (World Wide Web Consortium) provide different web metrics. Research work has been done to study web pages structures and the impact of Internet protocols on web browsing quality, but these solutions focus on a specific subject and do not analyze the whole chain of web browsing sessions (e.g., loading times, protocol distribution, types of services, location of web servers, etc.). Furthermore, those solutions make use of tools where JavaScript is sometimes disabled (whereas JavaScript is nowadays widely used in web pages), do not embark latest updated web browsers and are not connected to residential access networks when performing measurement campaigns leading to results qualifying web browsing quality which might be biased.

In this paper, we present Web View, a measurement platform which aims at overcoming such limitations by using probes connected to real residential access networks, launching real web browsers, and monitoring several parameters of the whole chain of Web browsing sessions. Since web pages' structures and used technologies regularly evolve, we make use of web browsers' network logs to obtain fine-grained information. Through our monitoring website, we visually represent the different loading times of a set of websites, the protocol distribution (assess their deployment rate), the number of downloaded resources and their web servers location, etc. In a second page, we show in detail the different web servers delivering content to end-users (origin or CDN servers together with their geographic location).

This paper is organized as follows: Section II describes the existing web metrics and related work meant to quantify web browsing, followed by Section III which illustrates our measurement platform. The section IV depicts our public visualization website and section V describes some remarkable events related to websites' evolutions and content delivery. We conclude in section VI.

## II. RELATED WORK

There have been several contributions from the research community to understand the Web ecosystem better.

Antoine Saverimoutou and Bertrand Mathieu are with Orange Labs, Lannion, France.
Sandrine Vaton is with IMT Altantique, Brest, France.

*Web metrics*. Aiming to bring uniform benchmarking indicators, standardization bodies such as the W3C together with large service companies and researchers have laid out a set of web metrics. The Page Load Time (PLT) is the time between navigation start until a web page is fully loaded. The Resource Timing provides some low-level information about the downloaded content (used transport protocol or type and size of objects). The Paint Timing exposes the First Paint (FP) which is the moment a first pixel appears in the web browser window. The Above-The-Fold (ATF) [3] exposes the time needed to load the visible part of a web page without scrolling by making use W3C web metrics. The Time for Full Visual Rendering (TFVR) [4] reflects the time to load the visible portion of a web page at first glance by making use of web browsers' networking logs.

*Web browsing quality*. Research works have been studying the impact of the transport protocols, the web page structures and the used technologies on web page loading times [5] and on end-users experience [6]. Research work has also been performed on different networking architectures which can improve content delivery by identifying the different objects to be served with higher priority by CDNs [7], exposing the different caching policies of CDNs [8] and identify the optimal transit selection from a CDN perspective [9]. Regarding web metrics, studies question their versatility and objectiveness and offer new techniques to improve their accuracy [10], new tools to better understand how web pages being wrongly designed can impact loading times [11] and how measurements setups can make web metrics to be biased [12]. Actual web metrics are mainly meant to measure static contents (dynamic contents are nowadays the de-facto standard).

*Measurement tools*. Different measurement tools have been used by the research community in order to perform automated web browsing. Among all these tools, OpenWPM [13] uses real web browsers to better quantify web privacy by supporting stateful and stateless measurements. Additional research work has also focused on web pages' structure and identified leading causes of degraded web browsing experience [14]. Eager to perform measurements on a larger geographic scope, web browser-plugins [15] have been proposed to measure in the wild the impacts of web page loadings.

This related work allows highlighting that each proposed solution is specific and that there is no complete user-representative tool investigating the whole Web browsing chain over long periods of time. These solutions do not embark various web browsers (and different versions) and are not attached to different home access networks (and different network operators). Our tool is regularly updated with new versions of web browsers and easily deployable on-the-fly everywhere. Compared to our platform, the other tools do not allow comparisons amongst multiple websites, neither aggregate different measurements nor allow an analysis into an historical way (e.g., comparing results from now versus values from 15 months ago).

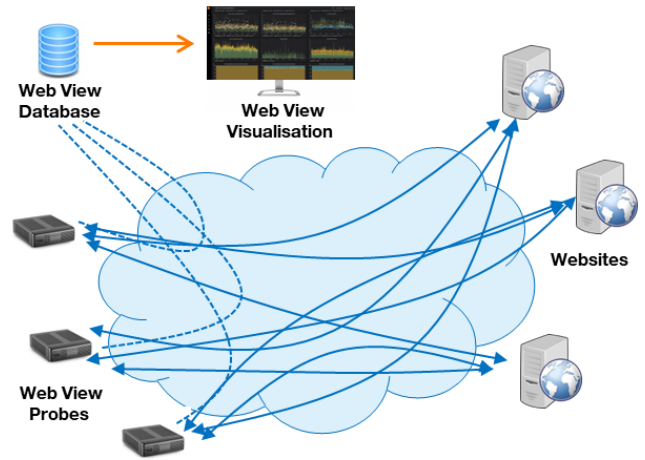

*Fig. 1: Web View infrastructure*

Web View embarks all web metrics available to date which helps in getting many information and link them so as to better understand the whole chain of web browsing sessions and the impact of each parameter. Finally, regarding content servers, the impact of the delivery via CDN, geographic location and corresponding protocol distribution is not really investigated in the research literature. The second functionality of our tool performs an analysis of this delivery part.

III. WEB VIEW PROBE

Web View is a measurement platform (Fig. 1), composed of 3 main components: the probes, the database and a public visualization website showing the obtained measurement results for a set of websites.

The Web View probe is a user-oriented measurement tool whose main objective is to perform automated web browsing sessions, measuring representative information of web pages in order to better qualify and understand web browsing, both in terms of performance and delivery. The probes emulate end-users' web browsing within a real end-user environment: real web browsers, residential access network, etc. Actually, 17 probes are currently deployed at 5 different geographic locations: 10 in Lannion (France), 4 in Paris (France), 1 in Vannes (France), 1 in Curepipe (Mauritius) and 1 in Tokyo (Japan). Our aim is to deploy more probes, worldwide, at various locations, in order to get many measurements results, from many places. For instance, we will install other probes in countries where Orange is, but deployment in other entities is also possible (as examples, we recently installed probes on other French network operators). With more probes we will be able to collect more data and analysis from the research community will help web developers to improve web page designs.

For each measurement test, we have to specify different configuration parameters: the web browser we want to use (currently Chrome web browser v.63-68-71-73-75-76-77 and the Mozilla Firefox web browser v.63-64-66-68-69 are supported), the access network of the probe since one probe is connected to only one access network (can currently be

```
{"name": "X-Cache",
 "value": "MISS, HIT"},
{"name": "X-App-Cache",
 "value": "HIT"},
{"name": "X-Served-By",
 "value": "cache-iad2132-IAD,
          cache-cdg20761-CDG"},
"serverIPAddress": "151.101.120.175"
```

*Fig. 2: Part of HAR file related to CDN delivery*

Fiber, ADSL or home WiFi of different network operators), the transport protocol we want to evaluate to get the contents (HTTP/1.1, HTTP/2, QUIC), the window size of the screen we want to emulate, the use of an ad blocker or not and the list of websites to measure.

It can be a list of pre-defined websites or the Top *N* Alexa websites (most visited websites on a daily basis: *https://www.alexa.com/siteinfo*). As we are performing many measurements all over the day/week/month, we scripted the different measurements tests to do and it is automatically launched via the script.

Since content servers might not implement all transport protocols, we allow fallback to protocols used by content servers. Typically, when performing measurements and requesting HTTP/1.1, we deactivate the HTTP/2 and QUIC protocol; requesting HTTP/2 implies deactivating QUIC but fallback to HTTP/1.1 is allowed; when requesting QUIC, we allow fallback to HTTP/1.1 and HTTP/2 for non-QUIC web servers; when requesting the Repeat mode (HTTP/2 or QUIC), we favor 0-RTT UDP and 1-RTT TCP connections by firstly navigating to the website, close the web browser, clear the resources' cache but keep the DNS (Domain Name System) cache. We then navigate once more to the website where measurements are collected. For every measurement, the resources' cache is always emptied and a timeout of 18 seconds is set to limit the impact of possible downtimes of content servers (value derived from all our measurements).

For every visited website, Web View probes offer 84 parameters. For the computation and collection of these parameters, we rely on the HTTP Archive (HAR) file which is the web browser's exposed networking logs. The use of networking information is privileged to W3C calculations in order to be more accurate when calculating loading times of dynamic and progressive content (Progressive Web Applications). For some parameters, we take the values offered by the HAR file, as others tools do, but for others, further processing tasks are performed leading to a richer information, than just what the HAR provides (e.g., the protocol distribution through which responses are delivered to end-users, the location of web servers, the different content providers, etc.).

Amongst the measured and computed 84 parameters by the Web View probes, we can mention 4 different loading times, namely the First Paint (FP), the Page Load Time (PLT), the

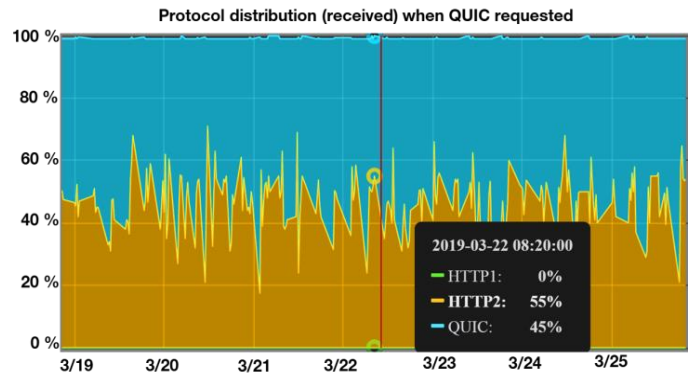

*Fig. 3: Protocol distribution for homepage "youtube.com"*

Time for Full Visual Rendering (TFVR) and the processing time, all obtained from HAR files.

The probe also provides information about resources composing the web page, i.e., the number of resources, their provenance, type, size and transfer rate. Since content can be delivered using a different transport protocol than the one we requested, we compute the distribution of received protocols for the given web page.

Finally, information about the content server is deeply processed in order to identify if a resource was provided by a content server or a CDN provider. For this, we use a part of the HAR file which indicates if the resource is retrieved from a cache (*HIT*) or not (*MISS*). For example, in Fig. 2, the first server replies with a *MISS* (cache-miss), meaning that the resource is not in its cache and the second server replies with a *HIT* (cache-hit) meaning that the needed resource is present in its cache. The corresponding content is thus delivered by the second CDN. From the exposed IP address, we perform a *WHOIS* to query the registered assignees, which results into the CDN *Fastly*. Relying on the MaxMind GeoIP2 database (*https://www.maxmind.com/*), we then identify the geographic location of the web server (town, country and continent).

## IV. WEB VIEW VISUALIZATION

The Web View visualization tool is a public website *(https://webview.orange.com)*, based on *Grafana*, and allows a straightforward visual analysis of our collected measurements. *Grafana* is connected to an *Elasticsearch* database, which stores all measurement results performed and sent by our Web View probes. The website offers several tabs illustrating how our platform works and has 2 main pages: one showing different panels related to the analysis of the web page browsing (loading times, resources, the protocol distribution, etc. ) and the second one representing content servers (CDNs or origin servers) on a world map, with information about resources and protocols.

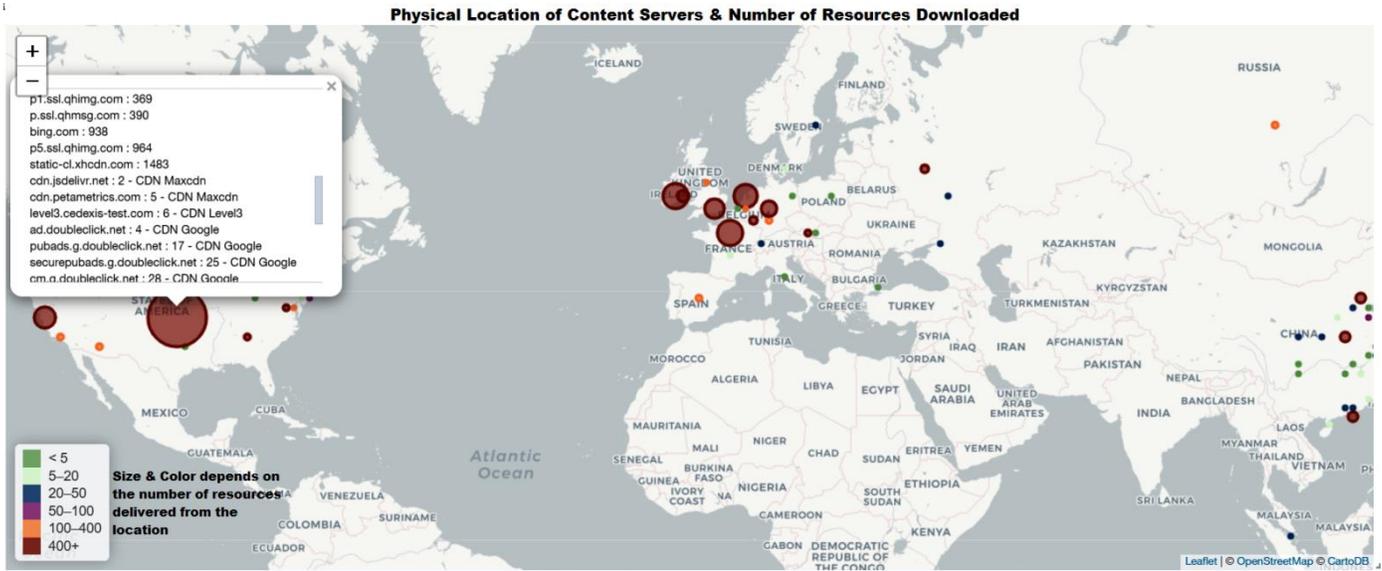

*Fig. 4: Location of origin web servers and CDN servers delivering resources to end-users in France*

For the web page offering websites' analysis, a menu allows a user to filter out several parameters: select a specific website, transport protocols, access networks, location of probes, web browser's window size, used web browser and use of an ad blocker or not. For the web page representing servers delivering contents, the menu offers an additional filter to select a specific CDN provider.

*A. Web Pages Analysis*

The first web page of our visualization tool depicts information to better understand the implications of the Web ecosystem during web browsing sessions. While some panels show the time needed to load a web page as a whole or through its progression by comparing different browsers (Chrome and Firefox), others compare the obtained timings when requesting specific transport protocols (HTTP/1.1, HTTP/2, QUIC). To better follow up with the deployment of the newest transport protocols, namely HTTP/2 and more recently QUIC, additional panels show the distribution of Internet protocols (how many resources are retrieved from web servers through HTTP/1.1, HTTP/2 or QUIC) as per a requested transport protocol. Additional information is also provided regarding the geographic location of web servers continent-wise. Following a set of filters selected by a user, all the panels are automatically updated.

For instance, Fig. 3 shows the protocol distribution when requesting QUIC and browsing the homepage of *youtube.com*. The web browser used is Google Chrome and we can see that even if Google promotes the QUIC protocol, on average 55% of contents are still distributed through HTTP/2 over TLS/TCP. When using an ad blocker, a mean number of 56 resources are downloaded from 8 different domains. These domains deliver all contents in a secured manner (HTTPS) and only 5 domains are QUIC-enabled.

*B. Content Servers Analysis*

The CDN web page allows to lay out figures on the delivery of contents, and mainly which web server provides a corresponding resource, the type of service delivered and their geographic location. An end-user located in Europe expects to firstly download objects from the same homepage domain and secondly from web servers located in Europe. Our visualization website allows to detect that at different times of the day, content might be fetched from different web servers at different geographic locations mainly due to Content Delivery Networks edge servers. Domains delivering contents and having an authoritative DNS name server different from the one of the homepage are entitled Non-Origin domains and conversely Same-Origin domains.

For instance, we can see through Fig. 4 that when an end-user is located in France and performs web browsing on the Top 50 Alexa websites, contents are mainly retrieved from France and Europe. Nevertheless, a non-negligible amount of content is also downloaded from North America and Asia. The different circles represent the distribution of content following the geographic location of web servers. Any user can hover the different circles and obtain fine-grained details about the web servers and the amount of content served. The Fig. 4 popup window illustrates the web servers delivering contents from a datacenter located in the Kansas County, United States.

Additional panels represent the Internet protocol distribution (HTTP/1.1, HTTP/2 and QUIC) from servers located all around the globe when performing web browsing measurements. Content servers referenced as *"No CDN"* imply that they are regular content web servers. From all our measurements performed on a set of websites, the list of CDN providers is automatically updated.

## V. REMARKABLE EVENTS

The Web View visualization website shows measurements performed since February 2018 where different web browsers have been used. Since web browsers are often updated (on average every month), the platform is also regularly updated. Measuring websites' loading times on large time-spans help in identifying the impact of actors of the Web ecosystem which can increase or decrease loading times. The Web View website being public, everyone can access it and analyze different behavior, but for people wanted to go into raw data, an open dataset can be found at: *https://webview.orange.com/public/WebViewOpenDataset.zip*. As examples, we present in this section some remarkable events we noticed over the past year.

### A. Web Browsers

Our tool makes use of two web browsers, Chrome and Firefox.

*Number of resources*. On average, 78% of our measured websites make use of Google services and when launching the Chrome browser, an average of 7 *Google objects* are downloaded and stored in the web browser's cache and used during further web navigation. When launching the Firefox web browser, no *Google objects* are downloaded which at the end of the day makes the Firefox web browser to download more objects in the PLT lapse of time.

*PLT timing*. Irrespective of the measured website, the Firefox web browser takes on average 1400 ms more time compared to Chrome to load an entire web page.

Over long time spans, we have been noticed that a higher loading time with Firefox web browser is mainly due to an increased processing time.

*TFVR timing*. When performing measurements with web browsers having a window size 1440 x 900, the Chrome version 75 loads the visible portion of a web page faster compared to Firefox version 64 since Chrome is embarked with a native ad blocker and downloads images in different formats of smaller sizes. The Firefox web browser TFVR timing is on average 750 ms more compared to Chrome. This points out that the different web browsers' engine policies can impact the perceived quality of an end-user.

*Processing time*. Chrome version 63 has been released in December 2017 and Chrome version 68 in July 2018. Chrome v.68 offers increased functionalities regarding the processing engine and security features. Firefox 63 has been released in June 2018. When comparing the different processing time of the above web browsers, Chrome 63 outperforms Firefox 63 on average by 900 ms, which plays a considerable role to the final exposed loading time (processing and network time). We have also identified that Chrome outperforms Firefox in terms of download and processing times since it downloads images in WebP format, being smaller in size, when Firefox downloads in JPG format (does not support WebP). It is important to measure processing time because, depending on the processing engine of the web browser, specific objects can be processed on a longer time and thus delay the download and processing of further downloaded resources.

### B. Transport protocols

Measurements are performed by requesting a specific transport protocol where fallback to a subsequent protocol is allowed if a distant web server does not implement it.

*Protocol distribution*. From our measured websites, when requesting HTTP/1.1, all web servers reply in strict HTTP/1.1 (any web server is HTTP/1.1 enabled which is a standard). When requesting HTTP/2, we can see that the overall HTTP/2 delivery has increased from 77% one year ago to 85%. When requesting the QUIC protocol, no website replies fully in QUIC following two reasons. Firstly due to the protocol mechanism itself, first requests to web servers are sent into HTTP/2 in order to assess if the corresponding web server is UDP-enabled. Secondly the QUIC protocol is mainly deployed on Google web servers. The QUIC Repeat mode can increase the QUIC distribution by up to 94%. As example, when considering the website *blogspot.com* (a Google website), requesting QUIC triggers the delivery of content through QUIC at a rate of 83% and QUIC Repeat enables the delivery of contents fully (100 %) in QUIC.

*Loading times following requested protocol*. When requesting HTTP/1.1, the average PLT of all our measured websites are on average 3400 ms. Requesting HTTP/2 reduces PLT values on average by 200 ms (not all web servers are HTTP/2-enabled). As indicated previously, the QUIC protocol is mainly deployed on Google web servers and requesting it creates a considerable fallback to the HTTP/2 protocol (UDP throttling) for non UDP-enabled web servers which increases on average the PLT by 100 ms. Requesting QUIC Repeat decreases the average PLT up to 1200 ms (compared to requesting HTTP/2). As an example, since May 2018, the website *amazon.com* has increased content delivery in HTTP/2 (from 92% to 98%) which has resulted in a decrease of 79% of PLT timings.

### C. Content servers

When browsing to a website, several domains from all over the world might serve contents to end-users. Fig. 5 shows the amount of objects downloaded from different continents between October 2018 and February 2019 when a European end-user browses the homepage *csdn.net* (tech blog). Until November 12 (Period 1), an average total number of 133 objects are downloaded when browsing to the website's homepage. On average 50 objects are retrieved from Asia, 2 from North America and 81 from Europe (CDN provider *Level 3*) through 63% HTTP/2 and 37% HTTP/1.1. Between November 13 and November 28 (Period 2), the homepage structure changes and a mean number of 123 objects are downloaded on average. These objects are served mainly from Asia (only 3 objects are downloaded from a CDN provider in Europe) which results in an increase of 3100 ms in the PLT. As from November 29 (Period 3), the CDN provider *Level 3* delivers on average 70 resources from Europe. When moving from Period 2 to Period 3, the average PLT is decreased by 32% thanks to the re-introduction of a CDN provider. When comparing Period 1 versus Period 3 where objects are delivered by a CDN provider, the average PLT is decreased by 36% mainly due to the shift in Level 3 edge domain web servers delivering contents in HTTP/2 (from 63% to 97%). Increasing the protocol distribution from HTTP/1.1 to HTTP/2 by 34% can reduce the PLT by 36% and up to 33% for the TFVR.

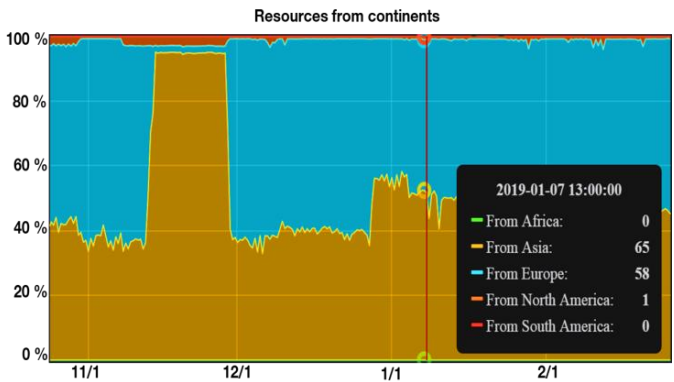

*Fig. 5: Location of servers for the website "csdn.net"*

| CDN Provider | Downloaded contents |
|---|---|
| Akamai | 60 |
| No CDN | 55 |
| Fastly | 25 |
| Google | 7 |
| Amazon | 2 |
| Cdn77 | 1 |
| KeyCdn | 1 |

*Table I: Content delivery for the website "lefigaro.fr" for an end-user located in Curepipe (Mauritius, Africa)*

With the CDN page of our website, we have been able to detect that the CDN provider EdgeCast, can now deliver some contents using the QUIC protocol, e.g., for the website *tumblr.com*. In May 2019, the QUIC protocol started to be officially discussed at the IETF (Internet Engineering Task Force) for standardization and we could notice in June 2019 that major CDNs (Fastly, Akamai, Verizon, etc.) started to implement the protocol. It allows us to follow the deployment of the QUIC protocol also in CDN products.

Websites referenced as *News* can make usage of several CDNs to deliver their content. As illustrated in Table I, when browsing to the homepage of *lefigaro.fr* (French news website) from Mauritius in the Indian Ocean, on average 151 objects are downloaded. From our visualization website, we can identify that 96 objects were downloaded from CDN providers and 55 objects from regular content web servers ("*No CDN*"). Among the different CDN providers, Akamai delivers the greatest amount of objects with respectively 45 objects from South-Africa, 12 objects from North-America and 3 objects from Mauritius (As shown in Fig. 2, objects may be cached by different CDN providers at specific geographic locations and the first CDN provider having the content in its cache is selected).

## VI. CONCLUSION AND FUTURE WORK

The Web browsing ecosystem is complex, involves several actors and resources are delivered by content servers located all around the world. In order to analyze it and identify which parameters play the most important role for web browsing, we proposed our Web View probe which makes use of real web browsers and residential access networks to measure finely web browsing sessions. The obtained measurements are visually represented, via our public Web View visualization website, so that any user can better understand the Web ecosystem.

The measurements performed by the probes for more than one year, allow us to better understand when, how and why the web browsing loading times might be increased or decreased: the used web browser, the transport protocol and the use of CDN impact them. Furthermore, having a history of measurements over more than one year, we have been able to see several evolutions of websites' content delivery. However, more probes deployed worldwide we can have, more measurements we can get and thus obtain more precise analysis and findings. In the next future, we aim at deploying probes in others countries where Orange operates network and opening the Web View platform to others entities could help to deploy probes in other locations and thus enrich the dataset of measured information.

In our technical future work, we will apply Deep Learning techniques such as Neural Networks Models (e.g., Random Forest) on the dataset to be able to predict loading times and non-parametric Bayesian Models (e.g., Hidden Markov Models) to capture patterns regarding resources delivery and loading times. This will allow to automatically detect changes in web browsing performance and delivery.

## VII. ACKNOWLEDGEMENT

This work is partially funded by the French ANR BottleNet project, No ANR-15-CE25-0013-001.

BIOGRAPHIES

**Antoine Saverimoutou** is a Ph.D student with Orange labs and IMT Atlantique (Brest, France). He obtained his Bachelor degree in Applied Mathematics at the University of Cambridge, Bachelor degree in Computer Science at the University of Caen and M.Sc degree from Caen, France. His thesis research is about Web browsing, and more specifically the study of new Internet protocols impact on end-users QoE.

**Dr Bertrand Mathieu** received a Diploma of Engineering in Toulon, the MsC degree from the University of Marseille, the PhD degree from the University Pierre et Marie Curie in Paris, and the « Habilitation à Diriger des Recherches » from the University of Rennes. He is working on programmable networks, QoS and QoE and new network solutions. He contributed to 11 national and European projects. He published more than 70 papers in international conferences, journals or books. He is member of several conferences Technical Program Committee and an IEEE senior member.

**Sandrine Vaton** is a full professor at IMT Atlantique (Brest, France). She has received an engineering degree and a Ph.D degree in signal processing from Telecom Paris, and then accreditation to supervise research (HDR) in computer science from Unity of Rennes. Her research studies are statistical modelling of telecommunications network and traffic, performance evaluation, network monitoring and anomaly detection.